\documentstyle[aps,prl,twocolumn,epsfig]{revtex}

\def\physd#1#2#3{{ Physica D}, {\bf #1}, #2 (#3)}
\def\prl#1#2#3{{ Phys. Rev. Lett.}, {\bf #1}, #2 (#3)}
\def\pla#1#2#3{{ Phys. Lett. A}, {\bf #1}, #2 (#3)}

\def\pre#1#2#3{{ Phys. Rev. E}, {\bf #1}, #2 (#3)}

\def\jsp#1#2#3{{ J. Stat. Phys.}, {\bf #1}, #2 (#3)}

\def\prc#1#2#3{{ Phys. Rep.}, {\bf #1}, #2 (#3)}

\def\eg{e. g.}

\def\beqr{\begin{eqnarray}}
\def\eqnr{\end{eqnarray}}
\def\beq{\begin{equation}}
\def\eqn{\end{equation}}

\def\bc{\begin{center}}
\def\ec{\end{center}}
\begin{document}
\title{Can Strange Nonchaotic Dynamics be induced through Stochastic
Driving?} 
\author{Awadhesh Prasad and Ramakrishna Ramaswamy}
\address{School of Physical Sciences, Jawaharlal Nehru University, New
Delhi 110 067, INDIA}
\maketitle
\begin{abstract}
Upon addition of noise, chaotic motion in low--dimensional dynamical
systems can sometimes be transformed into nonchaotic dynamics: namely,
the largest Lyapunov exponent can be made nonpositive. We study this
phenomenon in model systems with a view to understanding the
circumstances when such behaviour is possible. This technique for
inducing ``order'' through stochastic driving works by modifying the
invariant measure on the attractor: by appropriately increasing measure
on those portions of the attractor where the dynamics is contracting,
the overall dynamics can be made nonchaotic, however {\it not} a
strange nonchaotic attractor.  Alternately, by decreasing measure on
contracting regions, the largest Lyapunov exponent can be enhanced.  A
number of different chaos control and anticontrol techniques are known
to function on this paradigm.
\end{abstract}
\noindent
\section{Introduction}

The effect of noise on the dynamics of low--dimensional nonlinear 
systems has been widely studied.  One major motivation has been to 
verify the robustness of observed dynamical phenomena \cite{cfh,noise}, 
but a large number of studies are directed toward studying whether 
additive (or multiplicative) noise can induce novel dynamical phenomena.

In this context, noise--induced {\it ordering}  has been
extensively explored in the past few years
\cite{banavar,fh,jsp,longa,pikovsky}. At first glance, such results
appear counterintuitive since the addition of randomness would normally
be expected to enhance the effects of chaos in any system. At the same
time, it is well--established that additive noise causes phenomena such
as stochastic resonance \cite{sr}, or otherwise stabilizes chaotic
motion \cite{wacker}.  In other situations the effect is to reduce the
value of the largest Lyapunov exponent (LE), namely to make the system
less chaotic \cite{rajasekar}, or to create new random attractors
\cite{ashwin}.

Can strange nonchaotic attractors (SNAs) \cite{grebogi84,rmp} be formed
via stochastic driving of a nonlinear system? While it has been
suggested \cite{rajasekar} that additive noise can create SNAs, this
question touches upon an important open issue.  The only examples of
SNAs known to date have quasiperiodic driving in the dynamics
\cite{rmp}, although there are some experimentally studied systems
\cite{mandell,newexp} where the dynamics appears to be on strange
nonchaotic attractors, but where there is no explicit quasiperiodic
driving.  This question therefore has considerable practical relevance.

In this article we examine the mechanism whereby the addition of a
chaotic or stochastic signal to a general chaotic system has the effect
of reducing the degree of disorder \cite{rajasekar}. The particular
systems where this occurs all appear to have large contracting regions
in the phase space, typified by, say, an {\sl exponential tail} in the
Poincar\'e map. An additional motivation here is to understand the
different mechanisms \cite{fh,pikovsky} through which chaotic
attractors are ``made'' nonchaotic.  The contrast here is with
quasiperiodically driven chaotic dynamical systems which can often
transform a strange chaotic attractor into a strange nonchaotic one
\cite{grebogi84,rmp}. On strange
nonchaotic attractors (SNAs) the dynamics is aperiodic since the
attractor is fractal, but the largest Lyapunov exponent is nonpositive, so
there is no sensitivity to initial conditions. The dynamics is
intermediate between quasiperiodic and chaotic; there are features of
both regularity and chaos.

Our present results suggest that stochastic driving {\it alone} cannot
create SNAs: noise--induced stabilization differs in important respects
from strange nonchaotic dynamics. We find that the noise induced order
proceeds as follows.  By adding noise, the invariant measure on the
(noisy) attractor is modified. If the measure on those regions where
the dynamics is locally contracting is enhanced, then this has the
effect of lowering the Lyapunov exponent. (Alternately, the Lyapunov
exponent can be enhanced by increasing the measure in regions where the
local dynamics is expanding). On the other hand, given a
quasiperiodically driven system where there are strange nonchaotic
attractors, the addition of noise may also destroy such attractors
\cite{pre}.
 
Our main results are presented in Section II, where we discuss model
systems with stochastic forcing. We analyse the dynamics in terms of
local Lyapunov exponents \cite{pre,abarbanel} and show the methodology
of this mechanism for inducing ordering. A number of previously studied
control methods \cite{opf,mg,sp} appear to fall in this class of
techniques, as does a related anticontrol method \cite{ag}.  A summary
follows in Section III.

\section{Results}

Consider a stochastically driven nonlinear dynamical system 
specified by, say, the iterative mapping
\beq
x_{n+1} = f(x_n)+ \sigma \xi_n,
\label{eq1}
\eqn
where $\xi_n$ is additive stochastic or random noise of strength $\sigma$.
We consider the case when the system has positive 
Lyapunov exponent with no driving, {\it i.e.} for $\sigma = 0$. 
For nonzero $\sigma$ it can happen \cite{rajasekar} that 
the Lyapunov exponent corresponding to the $x$ degree of freedom
\beq
\label{lyap}
\lambda = \lim_{N \to \infty} {1 \over N} \sum_{i=1}^N \ln \vert 
f^{\prime}(x_i) \vert
\eqn
can become negative. 

A number of related situations show a very similar property, namely that
the Lyapunov exponent decreases on addition of an extra stochastic or
chaotic term in the dynamics. For example, driving via another chaotic 
system,
\begin{eqnarray}
x_{n+1} &=& f(x_n) + \sigma \xi_n\nonumber\\
\xi_{n+1} &=& g(\xi_n), 
\end{eqnarray}\noindent
where the maps $f$ and $g$ can be different from each other, or in an
extreme case, where $\xi_n$ is a constant, namely the case of constant
feedback studied in some detail by Parthasarathy and Sinha \cite{sp}.
This type of feedback causes the system, Eq.~(\ref{eq1}) to display a
form of ``control''. It may happen that a periodic orbit is stabilized
via feedback \cite{opf,mg,sp}. Alternately, the motion continues to be
aperiodic, but since the Lyapunov exponent is negative, two systems
with different initial conditions which are driven by the exactly same
noise will actually show synchronization.

However, in a trivial sense, the above dynamical system cannot possess
a {\it nonchaotic} attractor because the largest Lyapunov exponent,
namely that corresponding to the $\xi$ degree of freedom is positive.
If one considers two separate initial conditions, namely driving two
systems with independent realizations of the noise or chaotic driving,
then there is no synchronization, as can be expected.  In this feature,
such dynamics differs from the motion on strange nonchaotic attractors
where there can be robust synchronization \cite{syn}. One cannot have
true SNA dynamics in the presence of stochastic driving {\it alone}.

Some understanding of the above results can be obtained by considering
typical examples \cite{banavar,rajasekar}. For specific maps considered
the exponential logistic map,
\beq
\label{eq2}
x_{n+1} \equiv  F(x_n)=x_n \exp [ \alpha(1-x_n)].
\eqn
and the quadratic  logistic map,
\beq
\label{eq22}
x_{n+1} \equiv  F(x_n)= \alpha^{\prime}x_n(1-x_n), 
\eqn
These show the typical bifurcation diagram as a function of $\alpha$~
or ~$ \alpha^{\prime}$, with chaotic dynamics over a range in parameter
space.  In the latter case, Eq.~(\ref{eq22}), \eg~ at
$\alpha^{\prime}=4$, it is observed \cite{banavar,longa} that on adding
the restricted noise \cite{rnoise} the LE does not decrease: a
pair of such chaotic systems synchronize with identical driving noise
\cite{banavar}.
The logistic map also synchronizes with restricted noise, and in this
case, the Lyapunov exponent remains positive \cite{longa}.
\begin{figure}
\epsfig{bbllx=180pt,bblly=290pt,bburx=480pt,bbury=670pt,figure=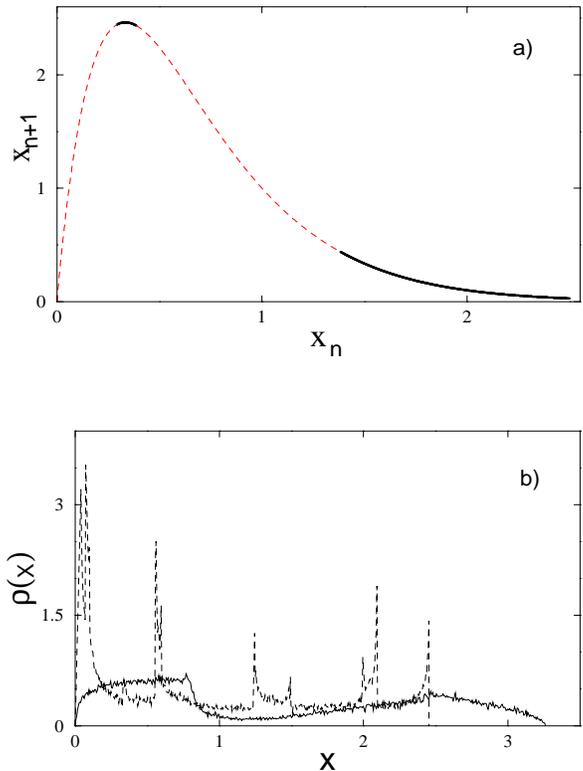,width=8.5cm}
\caption{
(a) The exponential logistic map with $\alpha=3$. Here, the dashed
lines are the expanding regions of the map, $\vert f^{\prime}(x)\vert >
1$, and the solid lines are the contracting regions.
(b) The invariant density $\rho(x)$ at $\alpha=3$ with noise strength
$\sigma = 0$ (dashed line) and $\sigma=0.1$ (solid line).  The noise is
$\delta$-correlated white noise.}
\label{fig1}
\end{figure}

Both maps have contracting and expanding sub--regions but
behave in different ways.  This difference can be analysed by
considering Eq.~(\ref{eq2}), for instance, at $\alpha=3$; the
dynamics is chaotic for almost every initial condition on $[0,
\infty]$.  Beyond $x \approx 1.4$ and in the region around the map
maximum [Fig.~1(a)], the map is contracting, so that only a small
region of the phase space is effectively responsible for the chaotic
dynamics. Most of the natural invariant measure is, however,
concentrated away from these contracting regions; see Fig.~1(b).
However for Eq.~(\ref{eq22}), the phase space is restricted only to
[0,1] and the contracting region is relatively much narrower.

The mapping Eq.~(\ref{eq2}) has positive Lyapunov exponent for $\alpha
= 3$.  By adding a noise term as in Eq.~(\ref{eq1}), the natural
measure can be modified so as to increase the sampling of the
contracting regions of phase space [Fig.~1(b)]: this reduces the
Lyapunov exponent of the driven system.  Consider the partial sums,
\beq
\label{sum1}
\lambda_+ = \lim_{N_+\to \infty} {1 \over N_+} \sum_{i} \ln \vert 
f^{\prime}(x_i) \vert,~~~~~~ \vert f^{\prime}(x_i) \vert > 1,  
\eqn
and
\beq
\label{sum2}
\lambda_- = - \lim_{N_- \to \infty} {1 \over N_-} \sum_{i} \ln \vert 
f^{\prime}(x_i) \vert, ~~~~~~ \vert f^{\prime}(x_i) \vert < 1,  
\eqn
namely the separate contributions to the Lyapunov exponent.  These are
obtained by partitioning a long trajectory ($N \to \infty$) into $N_+$
points on expanding regions and $N_-$ points on contracting regions.
Clearly, $N=N_++N_-$ and $\lambda=\frac{N_+}{N}\lambda_+-\frac{N_-}{N}
\lambda_-$. As the intensity of the noise term increases, the dynamics
is pushed out onto those parts of phase space where the average slope
of the map is less than 1. Thus the latter partial sum, $\lambda_-$,
increases in magnitude at the expense of $\lambda_+$ and eventually
becomes larger than $\lambda_+$, leading to a
Lyapunov exponent which is zero or negative.  
The variation of these quantities with noise
strength $\sigma$ is shown in Fig.~2, and it is clear that the system
can be made ``nonchaotic'' both in the case of additive noise, as well
as for driving via an added chaotic signal.

\begin{figure}
\epsfig{bbllx=30pt,bblly=80pt,bburx=520pt,bbury=520pt,figure=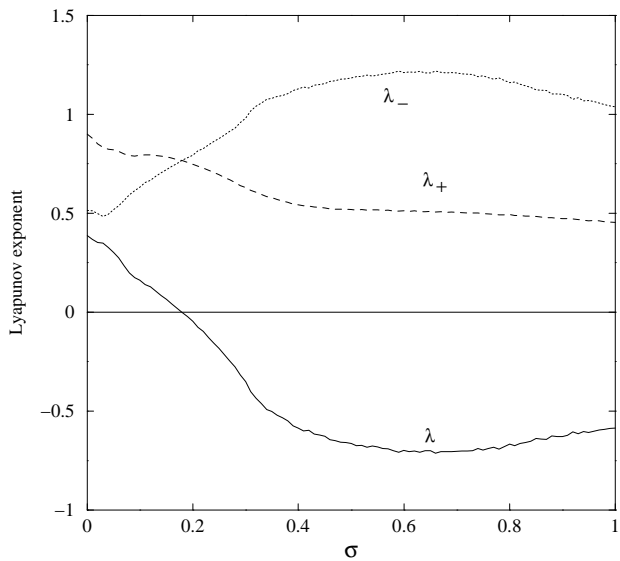,width=8.5cm}
\caption{ The Lyapunov exponent (solid line) and its partial sums,
$\lambda_+$ (dashed) and $\lambda_-$ (dots) [see Eqs.~(\ref{sum1}) and
(\ref{sum2}) respectively] as a function of the noise strength
$\sigma$.}
\label{fig2}
\end{figure}

Similar ideas apply to noise--driven flows where analogous results can
be obtained. Again, we find that systems where the Lyapunov exponents
can be reduced by adding noise are characterized by having a large
contracting region in the phase space; this can be detected by
examination of the return map, for instance. An important class of such
continuous system that we have studied are equations corresponding to
the kinetics of coupled chemical reactions, as for example the cubic
non-isothermal autocatalator (see Chapter 4 in Ref.~\cite{chem} and
references therein for more details).
\beqr
\frac{dx}{dt} &=& \mu \exp(z)-x y^2 -\kappa x\\
\nonumber
\frac{dy}{dt} &=& x y^2+\kappa x -y\\
\nonumber
\frac{dz}{dt} &=& \delta y-\gamma z
\label{chem}
\eqnr
\begin{figure}
\epsfig{bbllx=30pt,bblly=25pt,bburx=520pt,bbury=480pt,figure=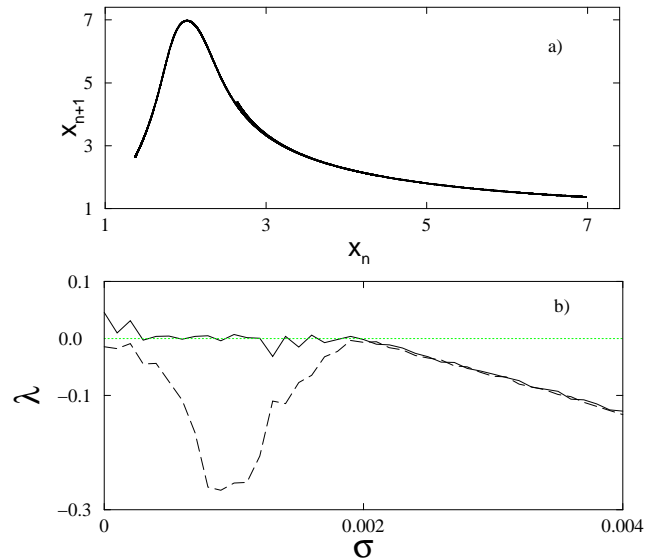,width=8.5cm}
\caption{ (a) Return map for the chemical reaction system,
Eq.~(\ref{chem}) with parameters $\mu=0.707, \kappa=0.00055,~
\delta=0.1$, ~and~ $ \gamma=0.5$  in the absence of noise; and (b)
the variation of the largest two Lyapunov exponents for
Eq.~(\ref{chem}) as a function of the strength of added noise,
$\sigma$.}
\label{fig3}
\end{figure}

Shown in Fig.~3(a) is the return map for the system in the regime where
the dynamics is chaotic, which shows an exponential tail similar to the
simple iterative mapping, Eq.~(\ref{eq2}).  Upon addition of noise the
Lyapunov exponents decreases as shown in Fig.~3(b).  Other examples
with very similar behaviour in higher dimensions are, for example,
equations that model the Belousov-Zhabotinsky reaction
\cite{chem,trap,bz}. Since these reactions can be studied
experimentally, it is possible that the effect of noise in reducing
chaos in such systems can be verified in practice \cite{trap}.

A number of different systems which share the above features
\cite{rajasekar,trap,hemmer} can be controlled in this manner, namely
by adding noise or chaotic driving. Note, however, that the system is
not truly ``nonchaotic''. Unlike the dynamics on periodic or
quasiperiodic attractors or SNAs where all the Lyapunov exponents are
nonpositive, here the dynamics is not confined to a single global
attractor, but to some region of the phase space for each realization
of noise. Therefore the fluctuations of all dynamical quantities, and
in particular the Lyapunov exponents, actually increase because of the
additive noise.

\section{Summary }

Nonuniform attractors in nonlinear dynamical systems typically have
interwoven contracting and expanding subregions.  By increasing the
measure on regions where the dynamics is locally contracting relative
to those which are locally unstable, one can render the motion
``nonchaotic''. This can be effected through the action of additive
noise: the invariant measure on some chaotic attractors can be so
modified \cite{jsp,rajasekar} that the dynamics is taken to those
regions of the attractor which are contracting on average, and this
results in a nonpositive Lyapunov exponent.  Indeed, noisy experimental
data can yield a negative value for the Lyapunov exponent even though
the actual dynamics of the system may be chaotic.

Adding stochastic noise, does not, however, create strange nonchaotic
attractors \cite{grebogi84}.  For each realization of the noise, the
limiting set is different, and thus there are no attractors {\it per
se}. One important property of SNAs is the synchronization of two
trajectories driven by the same external quasiperiodic force
\cite{syn}. Motion on the nonchaotic sets obtained by adding noise do
not have this property unless they are driven by {\it identical} noise.
Whether it is reasonable to expect that this can be realized in
practice is a moot question. 

In the present work, we have considered only additive stochastic
driving. It is possible that some other forms of stochastic driving
(perhaps via parametric modulation) can create true SNAs; this
question remains to be explored.

\vskip1cm
\centerline{\bf ACKNOWLEDGMENT} This research has been supported by a
grant from the Department of Science and Technology, India.


\begin{thebibliography}{99}
\bibitem{rnoise} The random numbers, $\xi_n$, are chosen such that the final
value, $x_{n+1}$ remains in $[0,1]$.
\end{thebibliography}
\end{document}